\documentclass[11pt]{article}
\usepackage{moriond,epsfig}
\def\be{\begin{equation}}
\def\ee{\end{equation}}
\def\bea{\begin{eqnarray}}
\def\eea{\end{eqnarray}}

\begin{document}
\title{THE DECAYS OF $B_c$ MESON TO $P$-WAVE CHARMONIUM $\chi_c(h_c)$ AND
TO LEPTONS WITH GLUON BREMSSTRAHLUNG\footnote{Presented in XXXVIIth Rencontres de
Moriond on QCD AND HIGH ENERGY HADRONIC INTERACTIONS, and 
in collaboration with Y.-Q. Chen, Z.-Y. Fang, A.K. Giri, R. Mohanta, 
G.-L. Wang, X.-G. Wu and H.-S. Zong}}
\author{Chao-Hsi CHANG (Zhao-Xi ZHANG)}

\address{Institute of Theoretical Physics, Chinese Academy of Sciences,\\
P.O. Box 2735, Beijing 100080, P.R. China}

\maketitle\abstracts{
Theoretical computations on the decays of $B_c$ meson to $P$-wave
charmonium $\chi_c$ or $h_c$ with some particle(s) else and on the decays
of $B_c$ meson to leptons with gluon bremsstrahlung, and the results as
well are outlined.
}

\section{Introduction}
\vspace*{-2mm}

The meson $B_c$ was observed successfully first by CDF collaboration in RUN-I
at Tevatron\cite{cdf}. CDF observation was through the cascade decay
$B_c\to J/\psi+\bar{l}+\nu_l$ with $J/\psi \to \mu+\bar{\mu}$ and the values of
$B_c$ mass, lifetime and a combined ratio of the cross sections and branching
ratios as below
$$m_{B_c}= 6.40\pm 0.39 \pm 0.13 GeV\;,\;\;\;\;
\tau_{B_c}=0.46 ^{+0.18}_{-0.16} \pm 0.03 ps$$
$$R\equiv \frac{\sigma(B_c)\cdot Br(B_c\to J/\psi l \nu)}
{\sigma(B)\cdot Br(B\to J/\psi l \nu)}=0.132^{+0.041}_{-0.037}\pm
0.031^{+0.032}_{-0.020}$$
are obtained.

Before the observation, in literature there were many theoretical estimates on
$B_c$ properties. Considering the theoretical uncertainties and experimental
errors, the previous theoretical estimates are consistent with the CDF observation.
Moreover, according to the theoretical estimates, experimental study of $B_c$ in LHC,
even in Tevatron RUN-II and RUN-III where numerous $B_c$ mesons can be produced,
is particularly interesting not only in understanding $B_c$ meson itself but also
for new physics studies, such as observing $CP$ violation in $B_s$, $B_s-\bar{B_s}$
mixing, and setting the more stringent constraint for multi Higgs doublet model
(MHDM, an extension of standard model with Higgs doublets more than two) etc.

The branching ratio of the decays $B_c\to B_s \cdots$ is very great
and the lifetime $\tau_{B_c} \simeq 0.5$ ps is very suitable for modern
vertex detector, so study of the $CP$ violation in $B_s$ and
$B_s-\bar{B_s}$ mixing in terms of the $B_s$ mesons from $B_c$ decays has great
advantages: the $B_s$ is tagged precisely ($B_c$ is charged) and many backgrounds
may be well rejected by the vertex of the $B_c$ decay. In MHDM the Feynman vertices
of charged Higges are proportional to the relevant Fermion masses, so the decay $B_c \to
\tau+\nu_\tau+\gamma$, especially, to measure $\tau$ transverse polarization, will
give much more tighter constraint than before on MHDM parameters\cite{india}.

Recently we completed computation on the decays of $B_c$ meson
to a $P$-wave charmonium $\chi_c$ or $h_c$ (semileptonic decays
$B_c\to \chi_c(h_c) \bar{l} \nu_l$ and nonleptonic decays $B_c\to
 \chi_c(h_c) h$, $h$ means a hadron) and obtained interesting results\cite{ccwz}:
the concerned decays are sizable (accessible in RUN-II, RUN-III and in LHC).
Especially, the decays to $h_c$ may be used as a new `window' to observe
$h_c$ state. The decays $B_c\to \chi_c \bar{l} \nu_l$ with $\chi_c\to
J/\psi \gamma$ followed, being a background, affect the observation of CDF
substantially etc.

A few years ago NRQCD made achievement in solving the puzzle
of $J/\psi$ and $\psi'$ production in hadron collision by the so-called color
octet mechanism, but it still has open problems. An outstanding problem is that
the experimental measurements on the polarization of the produced $J/\psi$ and
$\psi'$ in hadron collision are deviated from NRQCD predictions. Therefore to
explore the roles caused by the color octet components of the other `double heavy'
mesons such as $B_c$ meson, in addition to the charmonia, certainly is interesting.

According to NRQCD, $B_c$ meson state, similar to charmonia, should be an expansion
in Fock space:
\begin{equation}
|B_c>=O(v^{0})|(c\bar b)[{\bf 1}, ^{1}S_{0}]>+O(v)|(c\bar b)[{\bf 8}, ^{3}S_{1}] g>
+O(v)|(c\bar b)[{\bf 8}, ^{1}P_{1}] g>+...\;.
\label{eq:octex}
\end{equation}
To pursue the roles of the color octet components, we have computed the charged lepton
spectra and the width of the inclusive decay $B_c\to l+\nu_l+\cdots$ (here $`\cdots'$
means hadrons), and found that, indeed through measuring the spectra, the roles of the octet
components of $B_c$ may be verified experimentally\cite{ccfw}.

Here I only outline the topics mentioned above, whereas the details can be found in Ref. [3,4].

\vspace*{-2mm}
\section{The Decays of $B_c$ Meson to $P$-wave Charmonium}\label{subsec:prod}
\vspace*{-2mm}

To compute the semileptonic decays $B_c\to \chi_c(h_c) \bar{l} \nu_l$, all of the form factors,
related to the weak current matrix elements sandwiched by the states of $B_c$ and the $P$-wave
charmonium, should be calculated precisely. In the present case special care on the recoil
effects should be paid, because in the decays the recoil momentum can be great, even
relativistic. Thus the method, the so-called generalized instantaneous approximation
(GIA)\footnote{The method of the so-called generalized instantaneous approximation
was suggested first in Ref.[5].}, is adopted so as to treat the recoil effects properly.
As a result of the momentum recoil effects being treated properly, all of the form factors
(for the decays of $B_c$ to a $P$-wave charmonium $\chi_c$ or $h_c$) depend on
two `general functions' $\xi_1$ and $\xi_2$:

\begin{eqnarray}
({\epsilon}^{\lambda}\cdot\epsilon_0)\xi_1
\equiv\int\frac{d^3q'_{p'\perp}}{(2\pi)^3}
\psi'^{*}_{n1\lambda}(q'_{p'T})\psi_{n00}(q_{pT}),\;\;
\epsilon^{\alpha}_{\lambda}\xi_2\equiv\int\frac{d^3q'_{p'\perp}}{(2\pi)^3}
\psi'^{*}_{n1\lambda}(q'_{p'T})\psi_{n00}(q_{pT})
q'^{\alpha}_{p'\perp},
\label{q9}
\end{eqnarray}
where
\begin{eqnarray*}
&&d^{3}p_T=k^{2}_{pT}dk_{pT}ds d\phi\;,\;\;\;
q^{\mu}=q^{\mu}_{p\parallel}+q^{\mu}_{p\perp}\;,\;\;\;
q^{\mu}_{p\parallel}\equiv (p\cdot
q/M^{2}_{p})p^{\mu}\;,\;\;\; q^{\mu}_{p\perp}\equiv
q^{\mu}-q^{\mu}_{p\parallel}\;, \\
&&q_{p}=\frac{p\cdot q}{M_{p}}\;,\;\;\;
q_{pT}=\sqrt{q^{2}_{p}-q^{2}}=\sqrt{-q^{2}_{p\perp}}\;,\;\;\;
\epsilon_{0\mu}\equiv\frac{p_\mu-\frac{p\cdot
p'}{M'^2}p'_\mu}{\sqrt {\frac{(p\cdot p')^2}{M'^2}-M^2}}\;, \\
\end{eqnarray*}
$\phi$ is the azimuthal angle and $s=(k_{p}q_{p}-k\cdot q)/(k_{pT}q_{pT})$.
$\epsilon^{\lambda}_\alpha(p)$ is the $P$-wave (L=1)
orbital `polarization' in moving $p$ ($p^2=m^2$).

Based on GIA, the functions $\xi_1$ and $\xi_2$ may be calculated
numerically, as long as the wave functions of $B_c$ meson and the $P$-wave charmonium
$\chi_c$ and $h_c$ are calculated with potential model. The result is shown in Fig.1.
We may see from Fig.1 that $\xi_1$ may be negligible in the cases when momentum
recoil is tiny, i.e. it approaches to zero as the recoil approaches zero, and becomes
comparable with $\xi_2$ when the momentum recoil turns great. Here $\xi_2$ is `essential'
in a sense because it cannot turn to zero no matter how small (or great) the momentum
recoil is. Note that the present situation is different from the decays of $B_c$ to an
$S$-wave charmonium, where the one, being `essential' and similar to $\xi_1$, is always
dominant over the other (similar to $\xi_2$). Hence in leading order calculations, in the
case of $B_c$ to an $S$-wave charmonium, all of the form factors depend on $\xi_1$ only (all
the dependence on $\xi_2$ is ignored). The reason, why there is the difference, is that, in
the case of $B_c$ to a $P$-wave charmonium, the `essential' one ($\xi_2$) is suppressed, if
comparing with the normalization of the wave functions for $B_c$ or the $P$-wave charmonium,
so $\xi_1$ can become comparable with $\xi_2$ as the momentum recoil approaches great.
Whereas in the case of $B_c$ to an $S$-wave charmonium, there is no such suppression on
the `essential' one (the one similar to $\xi_1$) at all.

\begin{figure}
\centering
\includegraphics[width=1.6in]{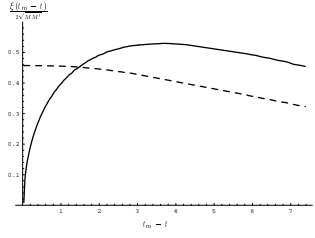}%
\caption{{\bf The functions $\xi_1$ and $\xi_2$ vs. $t_m-t$.} The solid line is
of $\xi_1$ and the dashed one is of $\xi_2$.}
\end{figure}

To give you some general feature, here I present the results of the
semiletonic decays only. The decay widths are put in Tab.1 and the spectra
of the charged leptons for
the decays are put in Fig.2 ($\tau$ lepton is massive, so its mass cannot be
ignored as done in the cases of $e,\mu$, thus Fig.2
contains two figures as described in the caption.)\footnote{Calculations on the
nonleptonic decays are straightforward as long as the effective Lagrangian including
QCD corrections and `factorization assumption' are adopted. The results can be found
in Ref.[3].}.

\begin{figure}
\centering
\includegraphics[width=1.6in]{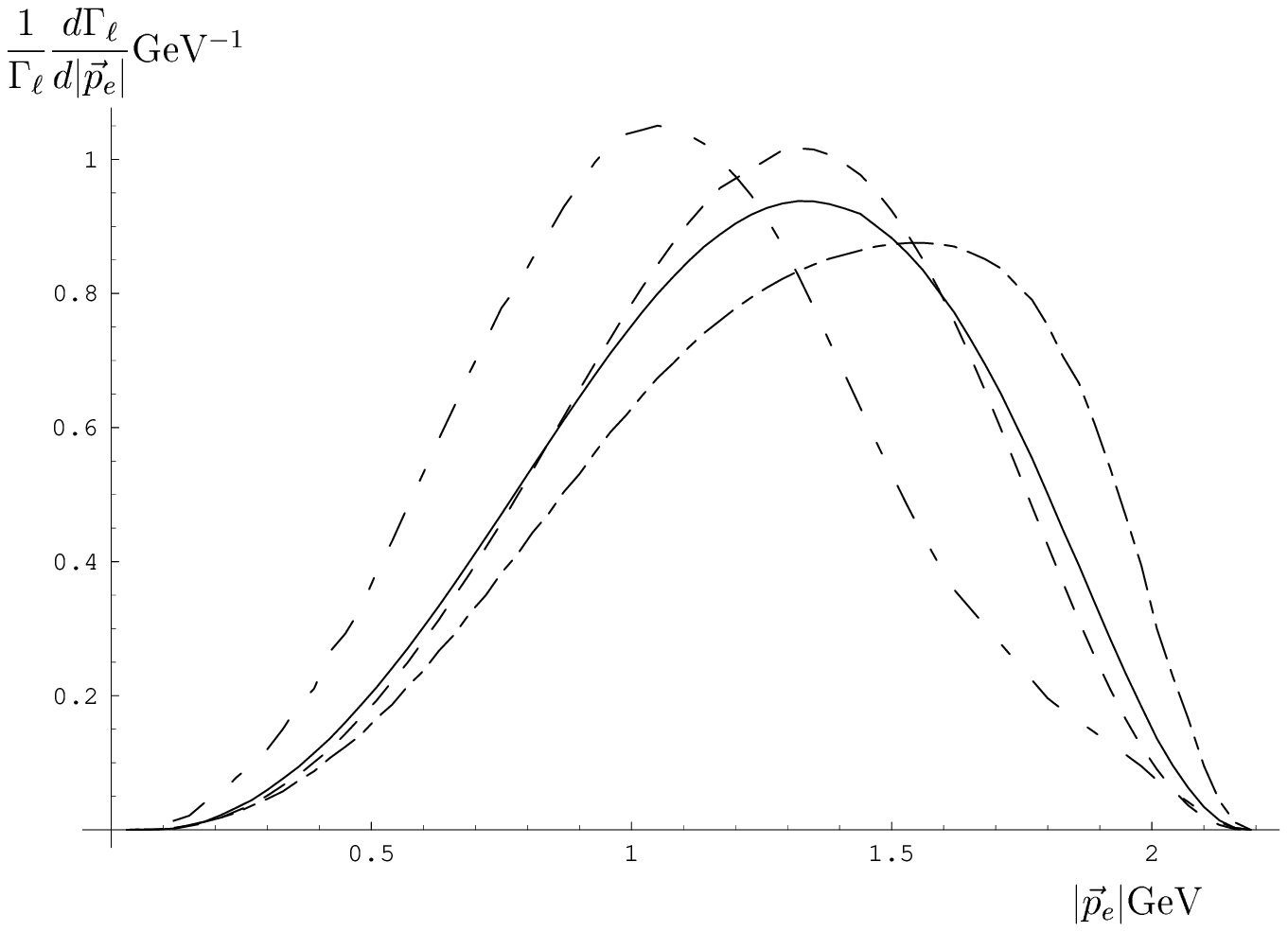}%
\hspace{0.8in}%
\includegraphics[width=1.6in]{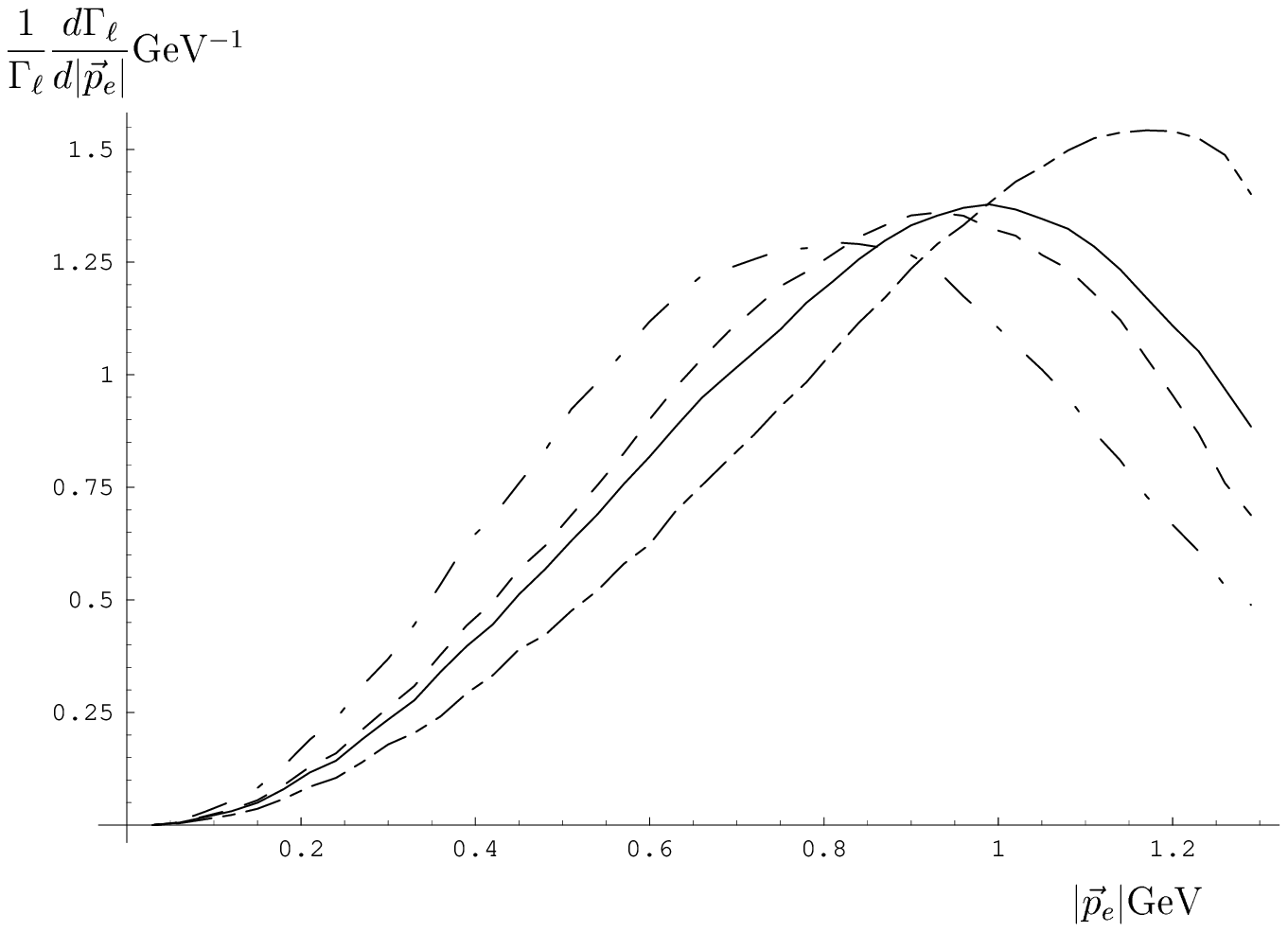}%
\caption{{\bf The energy spectra of the charged
lepton for the decays $B_c\rightarrow \chi_c(h_c)+e(\mu)+\nu_{e(\mu)}$ (the left figure)
and for the decays $B_c\rightarrow \chi_c(h_c)+\tau+\nu_{\tau}$ (the right figure).}
The solid line is of $h_c[^1P_1]$.
The dotted-blank-dashed line is of $\chi_c[^3P_0]$. The dashed line is of
$\chi_c[^3P_1]$. The dotted-dashed line is of $\chi_c[^3P_2]$.}
\vspace{-0.5cm}
\end{figure}

\begin{center}
{\bf Table 1: The semileptonic decay widths (in the unit $10^{-15}$ GeV)}
\vspace*{2mm}

\begin{tabular}{|c|c|c|c|c|}\hline\hline
&$\Gamma(B_{c}{\longrightarrow}{h_c}
{\ell}{ {\nu}}_{\ell})$&$\Gamma(B_{c}{\longrightarrow}{\chi_{c0}}
{\ell}{ {\nu}}_{\ell})$&$\Gamma(B_{c}{\longrightarrow}{\chi_{c1}}
{\ell}{ {\nu}}_{\ell})$&$\Gamma(B_{c}{\longrightarrow}{\chi_{c2}}
{\ell}{ {\nu}}_{\ell})$\\ \hline
$e(\mu)$&2.509&1.686&2.206&2.732\\\hline
$\tau $&0.356&0.249&0.346&0.422\\\hline\hline
\end{tabular}
\end{center}
One may see clearly from Fig.2 and the table that the branching ratios of the above semileptonic
decays (a lot of nonleptonic decays too\cite{ccwz}) are sizable if bearing $B_c$
lifetime $\tau_{B_c}\simeq 0.5$ ps in mind, especially, $B_c\to h_c{\ell}{ {\nu}}_{\ell}$,
can be used as a new window to observe $h_c$.

\vspace*{-2mm}
\section{The Decays of $B_c$ Meson to Leptons $l^+\nu_l$ with Gluon Bremsstrahlung}
\vspace*{-2mm}

There are two kinds of $B_c$ decays to leptons with gluon bremsstrahlung.
One is those induced by color singlet components of $B_c$ and the
other is those induced by color octet components of $B_c$.
In QCD the lowest order, the one induced by the color singlet components is
a decay with two gluon bremsstrahlung $B_c[{\bf 1}^1S_0]\to \bar{l}+\nu_l+g+g$,
whereas the ones induced by color octet components are
the decays with one gluon bremsstrahlung $B_c[{\bf 8}^3S_1]\to \bar{l}+\nu_l+g$
and $B_c[{\bf 8}^1P_1]\to \bar{l}+\nu_l+g$. The color octet ones are
of one order lower than that of the color singlet one, but based on NRQCD velocity
scale rule, the situation for the two kinds decays is inverse, so these two kinds
of processes may be comparable based on magnitude order consideration. Therefore
we computed them together and tried to see the possibility for experimental
verification of the color octet components\cite{ccfw}.

The Feynman diagrams relevant to color singlet one and the imaginary part
of Feynman diagrams relevant to the widths of the color octet ones
are shown in Fig.3. The final results of the calculated decay widths are proportional to
relevant matrix elements accordingly.

\begin{figure}
\vspace*{-2.8cm}
\begin{minipage}{0.45\textwidth}
\vspace{30mm}
\includegraphics[width=2.5in]{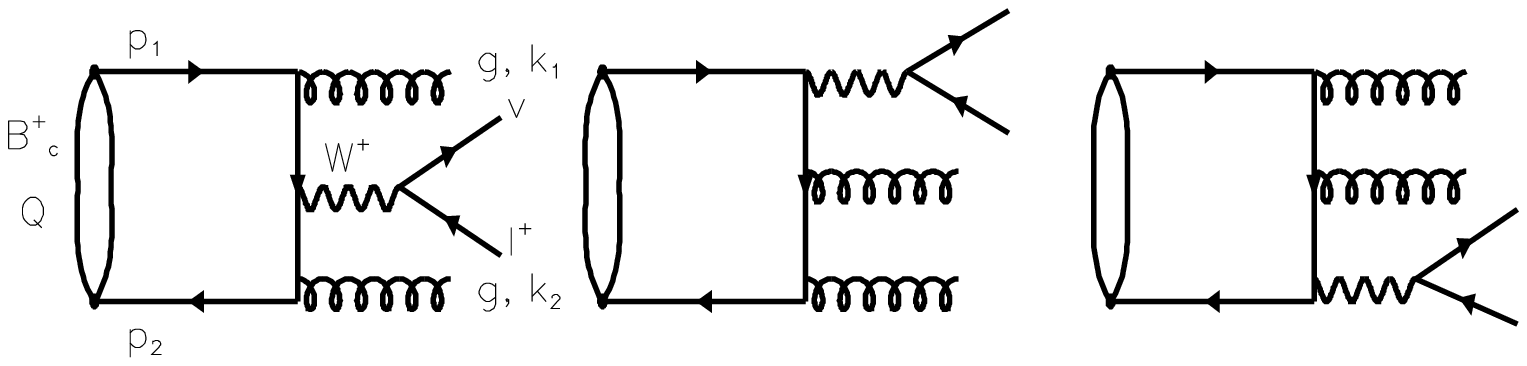}%
\end{minipage}
\hfill
\begin{minipage}{0.45\textwidth}
\includegraphics[width=2.5in]{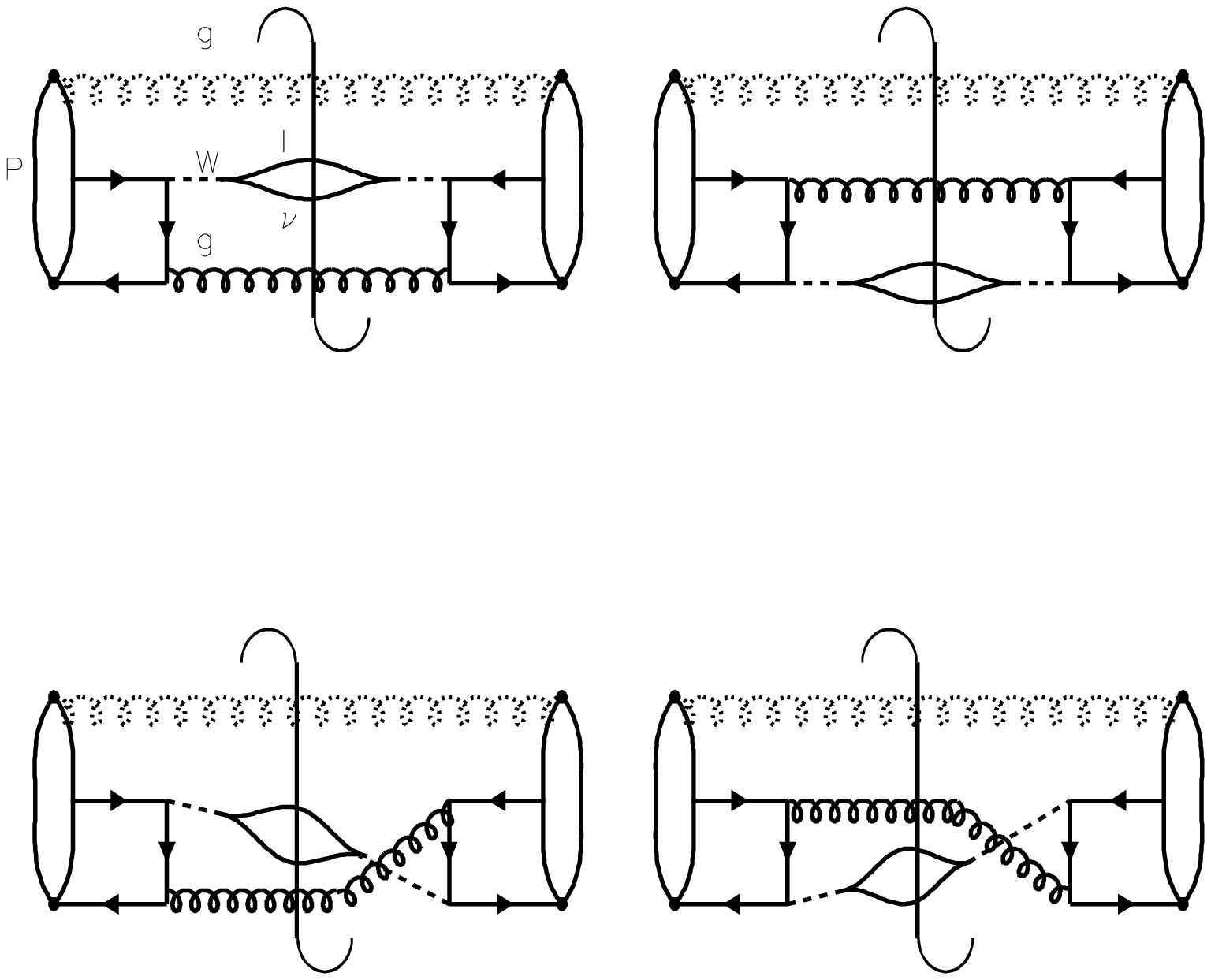}%
\end{minipage}
\vspace*{-30mm}
\caption{{\bf The Feynman diagrams:} left three Feynman diagrams are typical ones
for the decay $B_c\to \bar{l}\nu_lgg$ (there are six diagrams in total
for the decay and the other three can be obtained with
the two gluons `crossed'.). The right four diagrams are those
related to the widths for the decays $B_c[{\bf 8}^3S_1]\to \bar{l}\nu_lg$ and
$B_c[{\bf 8}^1P_1]\to \bar{l}\nu_lg$ respectively (Each of them has a
cut line and a dotted long gluon line. The former indicates to take imaginary part
of the diagrams and the later indicates a very soft gluon.)}
\vspace{-0.5cm}
\end{figure}

According to the velocity scale rule of NRQCD, the relative magnitude order of the
matrix elements for the color octet and color singlet components can be estimated.
Moreover the wave function at origin, relating to the matrix element for the color
singlet directly, can be computed by potential model, hence the
contributions of the decays concerned above to the directly measurable decay
$B_c\to \bar{l}\nu_l\cdots$ (here `$\cdots$' means hadrons) can be estimated
quantitatively. The numerical result for charged lepton spectra are put in Fig.4.
Since the velocity scale rule of NRQCD is only an order estimate, so for comparison
in the numerical calculations we tried two possible values for the color octet matrix
elements as explained in the figure caption.

\begin{figure}
\vspace{2mm}
\centering
\includegraphics[width=2.0in]{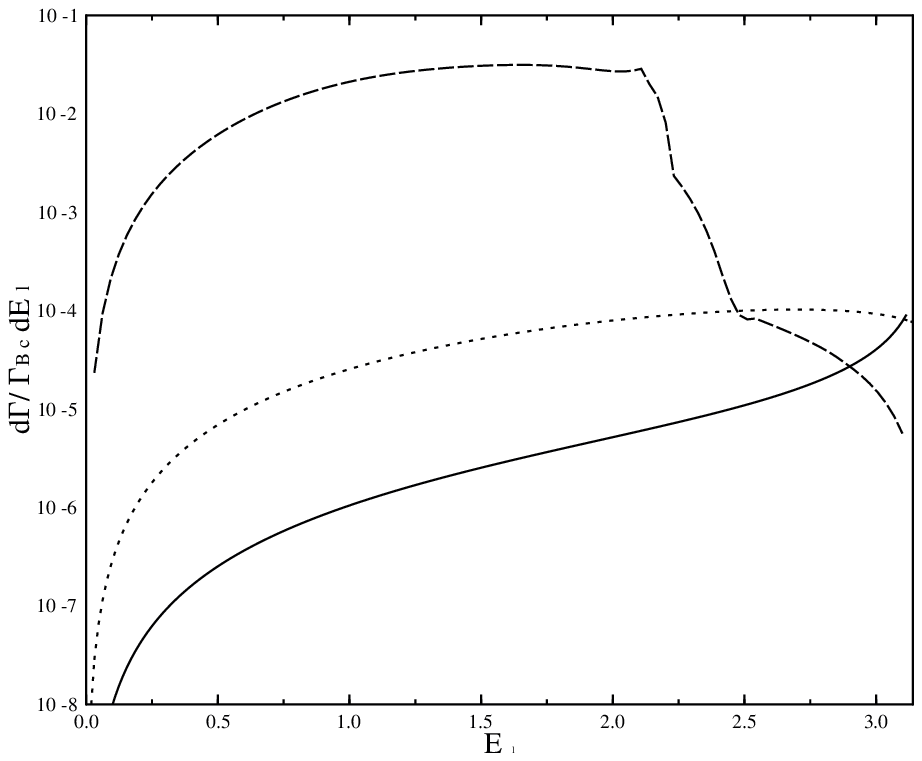}%
\hspace{1.0in}%
\includegraphics[width=2.0in]{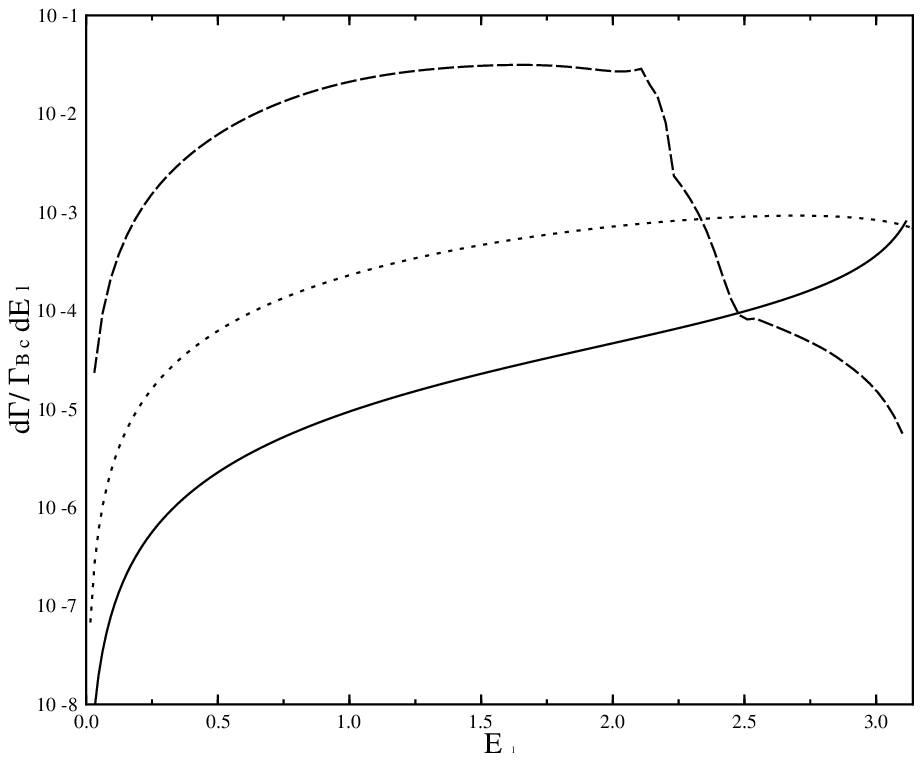}%
\vspace*{-6mm}
\caption{{\bf The energy spectra of the charged lepton for $B_c$ decays to leptons
with gluon bremsstrahlung.} The left figure is when color octet matrix
elements are taken to be ten percent of color singlet one.
The right figure is when the color octet matrix
elements are taken to be thirty percent of the color singlet one.
The dashed line in the figures stands is of the color singlet decay \(B_{c} \rightarrow
l\nu_l g g\). The dotted and the solid lines are of the color octet decay
\(B_{c} \rightarrow l\nu_l g \) with the \((c\bar{b}[{\bf 8}^{3}S_{1}])\)
and \((c\bar{b}[{\bf 8}^{1}P_{1}])\) components respectively.}
\vspace*{-0.5cm}
\end{figure}

It is interesting to see from Fig.4 that, in the region around the the end point of
the spectra of the charged lepton, the contributions from color octet components becomes
greater than those of color singlet, so that potentially it is possible to verify the
color octet components in $B_c$ meson through measuring the spectra carefully.

\vspace*{-3mm}
\section*{Acknowledgments}
\vspace*{-2mm}

I would like to thank Prof. J. Tran Thanh Van and Prof. E. Auge for invitation
and warm hospitality. The work was supported
in part by Nature Science Foundation of China (NSFC).

\end{document}